# Indigenous use of stellar scintillation to predict weather and seasonal change

Duane W. Hamacher [1,2], John Barsa [3], Segar Passi [3], and Alo Tapim [3]

[1] Monash Indigenous Studies Centre, Monash University, Clayton VIC 3800, Australia
[2] Mount Burnett Observatory, 420 Paternoster Rd, Mount Burnett VIC 3781, Australia
[3] Meriam Elder, Murray Island, QLD, 4875, Australia
Corresponding Author Email: duane.hamacher@gmail.com

**Abstract**

Indigenous peoples across the world observe the motions and positions of stars to develop seasonal calendars. Additionally, changing properties of stars, such as their brightness and colour, are also used for predicting weather. Combining archival studies with ethnographic fieldwork in Australia's Torres Strait, we explore the various ways Indigenous peoples utilise stellar scintillation (twinkling) as an indicator for predicting weather and seasonal change, discussing the scientific underpinnings of this knowledge. By observing subtle changes in the ways the stars twinkle, Meriam people gauge changing trade winds, approaching wet weather, and temperature changes. We then explore how the Northern Dene of Arctic North America utilise stellar scintillation to forecast weather.

## 1  Introduction

To the general public, the twinkling of stars serves as inspiration for art and poetry. While many Indigenous cultures draw their own poetic and aesthetic inspiration from this phenomenon, current ethnographic fieldwork shows that the twinkling of stars serves an important practical purpose to Indigenous peoples. A person's ability to accurately "read" the various changes in the properties of stars can assist them in predicting weather and seasonal change. This has application to navigation, time-reckoning, food economics, and predicting animal behaviour. The foundation for this knowledge has a scientific basis that has gone largely unnoticed or ignored by Western science. As Torres Strait Islander academic Martin Nakata told lead author Duane Hamacher, "We are not just a people of culture. We are also a people of science." This paper is one of many ongoing contributions that seeks to decolonise scientific discourse and acknowledge the rich and complex knowledge systems of the world's Indigenous peoples.

Since the development of cultural astronomy as an academic inter-discipline, our understanding of Indigenous Astronomical Knowledge Systems has rapidly expanded (e.g. Holbrook *et al.* 2008; Martin-Lopez 2011). The visible properties of stars, such as their color, brightness, changes in brightness, and relative positions with respect to the horizon all have special significance and applications to traditional Law and social structure (Hamacher & Norris 2011; Cannon 2014; Ruggles 2015). For example, stars can be grouped into asterisms, including dark asterisms traced out by the dust lanes in the Milky Way rather than stars. Particular asterisms or bright stars that rise or set at dusk or dawn foretell changing seasons and are often portrayed as important ancestral figures in Indigenous Creation stories (e.g. Cannon 2014; Hamacher 2012; Hamacher 2015).

The colors of stars often reflect the animals or ancestor-beings they represent. In Aboriginal traditions of the Great Victoria Desert, South Australia, ruddy Mars and the red-supergiant Antares both represent the red-tailed black cockatoo, and the red stars Aldebaran and Betelgeuse are associated with fire (Leaman *et al*. 2016). Narrindjeri traditions describe the variability of the pulsating giant Antares, which is linked to sexual passion (Hamacher 2018). Thus, observational and positional astronomy are an important and integral component of many Indigenous Knowledge Systems in Australia. But unlike Western science, Indigenous science is generally developed *in situ* (Verran 2002), meaning it is developed within the local environment and landscape in which the people live and interact (Nakata *et al*. 2014). This means the foundational knowledge may be the same across cultures separated by space and time, but the ways in which the knowledge is described in oral tradition or positioned in the Knowledge System can be quite different.

This paper analyses how stellar scintillation (twinkling) is understood and utilized by Indigenous peoples around the world, with a focus on the Strait Islanders of Australia as a case study. To show how this knowledge is developed *in situ*, we will also compare it with cultures in a vastly different environment, namely the Northern Dene of Arctic Canada and Alaska. The co-authors on this paper are Meriam elders who were interviewed for their knowledge, made significant contributions to the work, and reviewed the paper for publication. They are listed as part of an ongoing approach to recognising Traditional Knowledge custodians, their contributions, and their collaboration.

## 2 Stellar Scintillation in Global Indigenous Traditions

The twinkling of stars is a phenomenon described by Indigenous cultures around the world. For example, the Inuit traditions of Igloolik, Nunavut in Arctic Canada describe Sirius (Alpha Canis Majoris) as a 'dancing star', which they call *Kajuqtuq Tiriganniaglu* - the red and white Arctic fox (*Vulpes lagopus*) (MacDonald 1998: 75).[1] The perception of scintillation as 'dancing' is found across the world, including the Mocoví people in the Chaco region of South America (Martin-López 2013). In Wardaman Aboriginal traditions of northern Australia, scintillation represents stars 'talking to each other' (Cairns & Harney 2003: 22). Yup'ik custodian Bob Aloysius of Alaska describes the scintillation of the star Iralum Qimugtii (Sirius) as *Ingularturayuli*, meaning "one who always does Ingula dances" (Fienup-Riordan & Rearden 2013: 68).

A deeper look into these traditions reveals a wealth of practical knowledge about weather and seasonal change, something largely ignored in the scientific literature. In /Xam (San) traditions of Southern Africa, the heliacal rising of the stars Canopus (*!keisse*) and Sirius (*!kuttau*) denote the start of winter, which are seen to twinkle rapidly. This also coincides with an increase in westerly trade winds (which blow from the west), which peak in winter and drop in summer (Bleek & Lloyd 1911: 338-341). This is cause for a ritual ceremony where a /Xam man watches for the first appear-ance of Canopus in the morning (heliacal rising) in mid to late May, who then points a burning stick at the star. This welcomes the star, which represents a grandmother carrying rice, by providing her with warmth as she rises. In Yup'ik traditions of Alaska, the flickering of stars signals wind. If Sirius does not twinkle, fine weather is coming. But when it flickers quickly, a storm is approaching (Fienup-Riordan & Rearden 2013: 68). Bob Aloysius says that sometimes twinkling stars will appear to have small tails (Fienup-Riordan & Rearden 2013: 65-66). These tails indicate the direction of blowing wind. If the tails point south, the wind is coming from the north.

In the astronomical traditions of Oceanic cultures, wind is identified as the primary causal agent for stellar scintillation and is used to denote seasonal change. Hawai'ian navigators observe twinkling stars to predict weather and wind (Kamakau 1961: 36). Maori traditions of Aotearoa/New Zealand's South Island describe how the bright twinkling of Canopus (*Atutahi*) is caused by hard-blowing wind in October (Beattie 1990), signaling the approaching summer. In Samoan traditions, an unidentified twinkling red star (*Le Tāelo*) signals an approaching cyclone (Stair 1898). The people say the star has a "jagged appearance at the edges," indicating moisture and turbulence. Fijian traditions say that twinkling stars are indicative of northeast trade winds, which are important for sailing and navigation (Fison 1904: 69). Similarly, Wardaman Aboriginal people predict the wet season (October to April) from the bright twinkling of Canopus (*Ngilmungngilmung*) in the dawn skies of September (Cairns & Harney 2003: 142, 207).

Some Indigenous cultures, such as those of Australia's Torres Strait, distinguish planets from stars as the former do not (generally) twinkle (Haddon 1912: 219), a technique often utilised by Western astronomers. But Indigenous people do recognise that this is not always the case. The Euahlayi Aboriginal people of northern New South Wales note that when Venus is very low on the horizon, it can twinkle and change colour (Mathews 1981: 40). They describe Venus at those times as an old man making crude jokes and laughing animatedly (Fuller *et al.* 2014).

It is clear that Indigenous people around the world observe the twinkling of stars and incorporate this into their Knowledge Systems to forecast weather and predict seasonal change. However, very little has been published in the literature about the topic. To gain a deeper understanding of how stellar scintillation is utilised by Indigenous peoples, we explore an archival study across the Torres Strait and an ethnographic case-study with Meriam Islanders. The latter is based on ongoing fieldwork conducted by the lead author on Mer (Murray Island) in the eastern Torres Strait from 2015 to 2018 with key elders, some of who are listed as co-authors on this paper.

## 3     Torres Strait Islands, Australia

Astronomical knowledge is a central point of the sea-faring islander cultures of the Torres Strait, who inhabit the tropical archipelago between the tip of Queensland's Cape York Peninsula and the Papua New Guinea mainland (Sharp 1993). The Torres Strait is divided between two major languages: Meriam Mir - a Papuan language spoken in the eastern islands, and Kalau Lagau Ya (and various dialects) - an Aboriginal-linked language spoken in the remaining islands. Torres Strait cultures share close connections between the Aboriginal people of mainland Australia and the Melanesian people of Papua New Guinea. The Kaurareg people of the southwestern island group, identify as Aboriginal rather than Torres Strait Islander.

Islanders observe the Sun, Moon, and stars to determine when to fish, plant and harvest gardens, hunt turtle and dugong, and the changing of seasons (see Eseli 1998; Haddon 1912; Sharp 1993). George Passi (1986: 66), a Meriam academic from Dauar, explains:

> *The [Islanders] also know the right seasons to engage in such [preparation] activities by observing the behavior of the stars and constellations in the sky. In fact, it is the stars or constellations that are seen to govern the behavior of plants and animals, which in turn influences the subsistence activities of the Islanders. The stars and constellations act as important guides to such activities.*

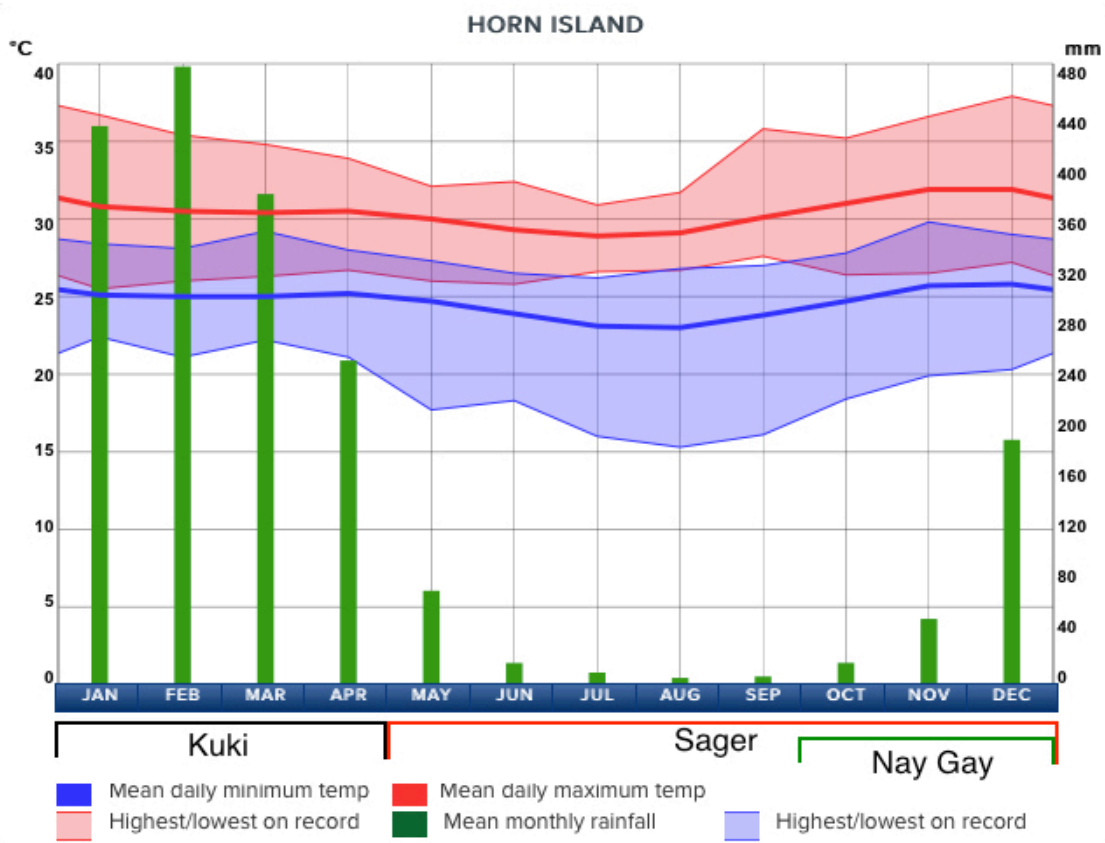

***Figure 1****: Long-term climate data for Ngurupai (Horn Island,) Torres Strait (1995-2015). Image from Weather Zone, URL: www.weatherzone.com.au/qld/peninsula/murray-island. Islander season names added by Hamacher.*

To understand the role of stellar scintillation in predicting weather and seasonal change, it is important to understand the climate of the region. Seasons in the Torres Strait are generally guided by trade winds and rainfall, with little change in temperature and length of day throughout the year (Figures 1 and 2). Islanders denote four general seasons throughout the year (with some overlap):

- **January to April**: *Kuki* (wet) season. It is a period dominated by strong northwest trade winds and monsoons.

- **May to December:** *Sager* (dry) season. It is a period dominated by southeast trade winds and dry weather.
- **October to December:** *Nay Gay* (hot/humid) season. It is a period of northerly trade winds and high temperatures and humidity.
- **Throughout the year**: *Zey* – southerly winds that blow randomly.

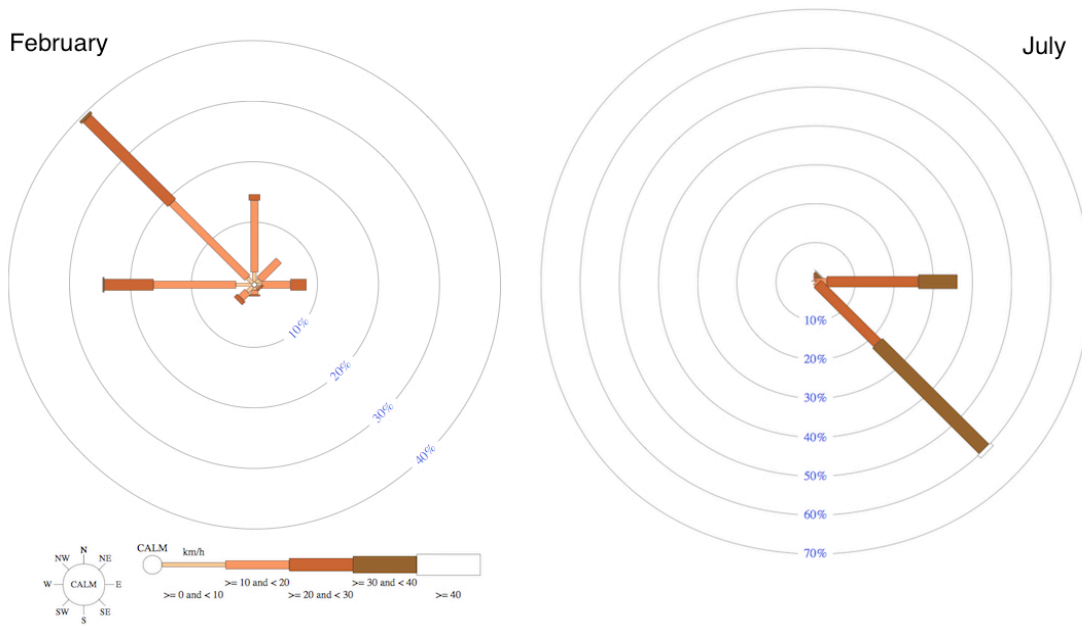

*Figure 2:* Rose of wind direction vs wind speed, measured at 3:00 PM daily from Horn Island between 1995 and 2010 (based on 415 total observations). The rose shows that February (mid-Kuki) is dominated by northwesterly winds while July (mid-Sager) is dominated by southeasterly winds. Australian Bureau of Meteorology.

A common concept described by elders during interviews regards one's ability to "read the stars". This is in reference to observing and interpreting subtle changes in the properties of stars. For example, George Passi (1986: 51-52) interviewed a Saibai man who told him that:

> *The moon and stars tell the hunters the kind of weather they will be expecting the next day. The lightning and the formation of the clouds will tell from what direction the wind will be blowing, how it will change its course later that day and where it will blow until sunset.*

The late Meriam man and co-author John Barsa explains how stellar scintillation and the changing colours of twinkling stars inform people about winds, coming storms, and fine (hot) weather. Blue/white twinkling stars that rapidly change colour to red/orange indicate wind and variations in temperature. Blue/white twinkling stars that do not change colour indicate fine weather and constant (hot) temperatures, which is common during the *Nay Gay* season (November to December). Blue/white stars twinkling rapidly indicate approaching storms and are common near the start of the *Kuki* (monsoon) season (January to April). The rapidly twinkling stars during this time are caused by strong northwest trade winds.

An elder (who chose to remain anonymous) demonstrates how Meriam people determine if twinkling stars are caused primarily by wind movement alone, or wind movement in conjunction with temperature variations and moisture. The elder explains that when stars "twinkle hard" (rapidly) and the wind drops, it is a good time to go fishing. If fishermen are out on the reefs, they can read the stars to know if they are safe to travel back home or if they should camp overnight on the reef. If they see stars twinkling in such a way that indicates wind movement, they know they are ok to head back.

Elder and co-author Alo Tapim explains how Meriam fishers and gardeners observe their surroundings, including plants and the stars, to know when the Kuki is coming. At the end of the hot, dry season (*Nay Gay*), the sky is very clear and the stars shine brightly. When they begin to twinkle, it signals increasing winds. Tapim explains:

> *There are various signs that the seasons change. We see it in trees and gardens, but mainly in the stars. It becomes quite apparent towards the end of the year when you get the very clear skies and hot days because there are no clouds, and the same through the night. And the stars are there very clear and colourful. When they twinkle, it heralds the wind picking up and fishermen capitalize on that – giving them the idea that they will have a good day at sea. Or they can overnight at the reef and come back the next day by observing these stars. From spring to summer, we get that dry, hot season where there are hardly any clouds in the sky. The word is doldrum, very fine – not a breath of air – it's very still. During the night, when the dew sets in, you see the stars. They're really shining and twinkling. It's a good sign. It heralds the changes in the weather. Instead of the hot dry season, a wet season is coming. Locals always look forward to that. They observe the stars because the gardener is anxious. When is that rain going to come? They observe the stars, the plants, and the path of the sun. They are aware there is dew at night, even though the sun is hot during the day. The dew rolls right into the dry ground, even though the ground may look brittle on top. When a farmer digs down, it is cool and muddy inside, and this is a result of the dew. Dew keeps the soil soft inside. Maybe a baby cicada or something not fully developed is present. By observing that cicada, it is evident the Wet [season] is near. So by observation of the ground, the environment, and the sky at night, the farmer is reassured that the wet, rainy season is coming.*

Erub elders explained to McNamara *et al.* (2010: 7) that the doldrum is very important and seen as a separate season by many elders. It is couched in the roughly three-month interval between the dry and wet seasons, during the *Nay Gay* season from October to December.

Mabuiag elder Peter Eseli (1886-1958) told Matsumoto (1983: 359) that Islanders link heavy winds, temperature changes, and coming rains to varying degrees of stellar scintillation. Towards the second half of the dry season, northeast winds bring rain. The end of the dry season brings gusty winds and heavy squalls, the temperature drops to its lowest point, and rain clouds from the northwest bring the monsoons. During this time, Vega and Delphinus are "active in the heavens" (a euphemism for scintillating). In Mabuiag traditions (Eseli 1998), the stars of the Western constella-tion Delphinus are called *Buu*, the trumpet shell, and the star Vega (Alpha Lyrae) is *Dhogay Ii*. The Kuki season experiences strong, tearing northwest winds, and is foretold by *Dhogay Ii's* brief appearance in the dusk sky before setting, and its heliacal (dawn) rise during the height of *Kuki* in late January.

Meriam elder and co-author Segar Passi explains that stars twinkle more under the influence of high altitude trade winds. He said the stars are brighter and twinkle more towards the end of the year.[1] The rapid twinkling of the stars during this time signals arrival of the northwestern *Au Uag* ("big wind") of the *Kuki*. This is supported by an unnamed Erub elder, who told McNamara *et al.* (2010: 7) that twinkling stars signal the start the northwest winds:

> *When they start to get the little rollers on the beach, it's flat calm, and they watch the stars, when the top stars, they twinkle more quickly, when the others more slowly it tells you something. Anyway, he read all the signs and said, 'It's gonna blow shortly', and then they start to get the rollers… Biru Biru is the one that flies out. They [birds] are signs of the nor-west [wind].*

Meriam knowledge about stellar scintillation and trade winds at the start of the *Kuki* are encoded song. A song composed by the late Meriam man and academic George Passi describes the affect the northwest trade winds have on the twinkling of stars. The song (with lyrics in the Meriam Mir language) describes a calm, clear night with stars twinkling like embers in a fire. The cause of this the twinkling is the "big wind", bringing rain clouds as winds shift from the dry southeasterlies to the wet northwesterlies.

It is obvious to the lead author that this song is known to most Meriam people, as evidenced by the many times the song was sang collectively by children and adults during festive events and gatherings. It is one of the primary ways the foundational knowledge of the phenomenon is passed to successive generations. A number of Meriam people acknowledged to the lead author that they watch for the twinkling stars to get an idea of the weather for fishing and hunting, which are still common practices. The knowledge is current, thriving, and used daily.

## 4     Northern Dene of Arctic North America

Ethnographers are working with Indigenous communities in other parts of the world to record similar knowledge of the phenomenon. University of Alaska anthropologist Christopher Cannon works closely with Northern Dene communities in Canada and Alaska. His Master's thesis (Cannon 2014) shows that these communities utilise star twinkling in similar ways as the Torres Strait Islander, but with specific application to their inland, Arctic environment *in situ*.

Much like the Torres Strait Islanders, astronomical knowledge is a fundamental component of Northern Dene/Athabaskan traditions in Arctic North America (Cannon 2014: 2; Cannon & Holton 2014: 5). The Northern Dene utilise stellar scintillation to facilitate navigation, time-reckoning, religion, and cosmology (Cannon 2014). The stars are observed as one of many natural signs to forecast the weather, and how the people read stellar scintillation is an important component of this knowledge.

---

[1] In this context, the elders are referring to the end of the year as December, not in an Indigenous context.

Ahtna people of south-central Alaska use the phrase *son' neke nałts'iihwdelae* ('the wind is sweeping the stars') to describe brightly twinkling stars that forecast a strong wind. Similar to the relationship between wind and scintillation described by the Meriam people of the Torres Strait, an Ahtna elder explained to Cannon (2014: 121) that :

> *We know wind blew too. Son' ['star'] it start moving all over. Big wind is headed down here pretty soon… Son' you know. Son' neke nałts'iihwdelae. This world pretty soon going to be big wind down here. We know that, you know. Son' neke nałts'iihwdelae, just like he sweep over. Not hanging in one place, you know; everyplace.*

In Ahtna language, the phrase *tloon' naa'ełts'eeyhleyaah* ('the wind swings the stars') similarly denotes brightly twinkling stars that indicate a strong wind (Jette *et al.* 2000: 585). Alternatively, blurred stars are called *tloon' tl'ołeł yee daadletl'ee* ('the stars are sitting in their cradles'). Dene elders across the Arctic describe twinkling stars in terms of dancing, including the Sun (although this is caused by a different phenomenon to scintillation). In the Great Slave Lake region, the "dancing" of the Sun (*sa datło*) is caused by rising heat during the warming weather in March and April (Cannon, 2014). The Gwich'in and North Slavey people observe the Sun appearing to "bob up and down" in the mountains during April (Cass *ca.* 1959: 6). This phenomenon was used, along with other indicators, as a tool in predicting local temperatures.

Changes in local temperature are also predicted by observing the intensity of stellar scintillation. In Gwich'in traditions of Fort Yukon, the stars appeared more 'magnified' in September before the arrival of the intense cold (Cannon 2014). Elders note that when the weather is colder, the stellar scintillation is less prominent (described as 'little twinkles') than in warmer weather, when scintillation is more pronounced ('larger' and 'bright' winkles). Northern Dene observations of stellar properties (in conjunction with weather forecasting) are often described in the context of cross-country travel and subsistence actives and that occur between September and April.

## 5      Underpinnings of Indigenous Knowledge of Scintillation

Indigenous Knowledge Systems have a scientific underpinning that is derived through empirical observation, experimentation, and deductive reasoning (Mazzocchi 2006; Nakata 2010). This knowledge is often passed to successive generations not through written word, but through oral tradition. Oral traditions are passed through strict protocols to ensure vitality and longevity, primarily utilising the *method of loci* (Kelly 2015). Indigenous Knowledge with respect to stellar scintillation exemplifies how Indigenous people closely observe subtle changes in the properties of stars and assign purpose to that knowledge.

Here, we explore the ontological crossroads of Indigenous Knowledge and Western Science. This is a contested space, an area encapsulating what Nakata (2002, 2007, 2010) calls the *Cultural Interface*. Although Indigenous and Western scientific con-cepts of what we might consider "natural phenomena" have different understandings and applications, there is a crossroads at the foundational level. In a positivist sense, we can learn how 'Nature' and the Universe work – at least to a degree. How that knowledge is produced, conceptualised, encoded in tradition (be it story and song, or textbooks), or applied to daily life can differ widely between Indigenous and Western perspectives. But there is a space where these two knowledge systems intersect. It is this space that we explore. We see that Indigenous people and Western science share a fundamental understanding of stellar scintillation.

In this paper, we explored how different Indigenous peoples conceptualise and apply knowledge of stellar scintillation, with a focus on the Torres Strait. We now explore Western theoretical foundations of stellar scintillation using the physics of electromagnetic scintillation, as reviewed by Sterken & Manfroid (2012). To the naked eye, stars appear as point-sources of light that are heavily affected by changes in the Earth's atmospheric conditions. In scientific terms, these changes are brought on by a combination of turbulence and variations in air density, humidity, and temperature (Sofieva *et al.* 2012). Turbulence produces moving pockets of air with varying densities that alter the refractive index of incoming starlight (Aerts *et al.* 2010). Wind velocity is proportional to the rapidity of scintillation. Thus, the faster these air pockets move, the faster the stars scintillate – as noted by the various Indigenous knowledge custodians across the world. Scintillation is the rapid change in stellar brightness due to atmospheric effects (but not blurring or changes in colour). Astronomers use the term "astronomical seeing" to describe these combined effects.

Starlight passing through a thicker layer of atmosphere means that the light is more susceptible to seeing effects. Brighter stars are also less susceptible to this than fainter stars, as their light flux is greater. This is consistent with Meriam elders saying it is important to look at high-altitudes ("top stars") to best interpret seeing effects and their relationship to wind velocity, as opposed to stars lower on the horizon that are passing through a thicker layer of atmosphere and are thus more affected by the atmosphere. It also explains why Sirius (Figure 3) and Canopus (Figure 4) are the named stars in the traditions described in this paper. These are the two brightest stars in the sky and are ideal for observing the prominent effects of scintillation. As point sources of light, stars are more affected by scintillation than planets. Planets are close compared to stars and their incoming flux is substantially higher. Therefore, planets tend to only twinkle when they are very low on the horizon or are observed through an extremely turbulent atmosphere. This is the foundation of the Euahlayi tradition about the man represented by Venus twinkling rapidly, which represents the man laughing when the star is low on the horizon.

In both Indigenous and Western scientific terms, wind movement (turbulence) is the primary cause of scintillation. Wind speed and the speed of scintillation are proportional: slowly twinkling stars denote slow winds while rapidly twinkling stars denote strong winds. Strongly scintillating stars with a jagged appearance at the edges signal a coming cyclone, denoting pressure, temperature, moisture, and wind speed variations that blur and distort the 'sharpness' of the star to our eyes. Elders in both the Torres Strait and the Arctic demonstrated awareness that air high in the atmosphere can be quite turbulent, causing the stars to twinkle even if it is calm at the ground level.

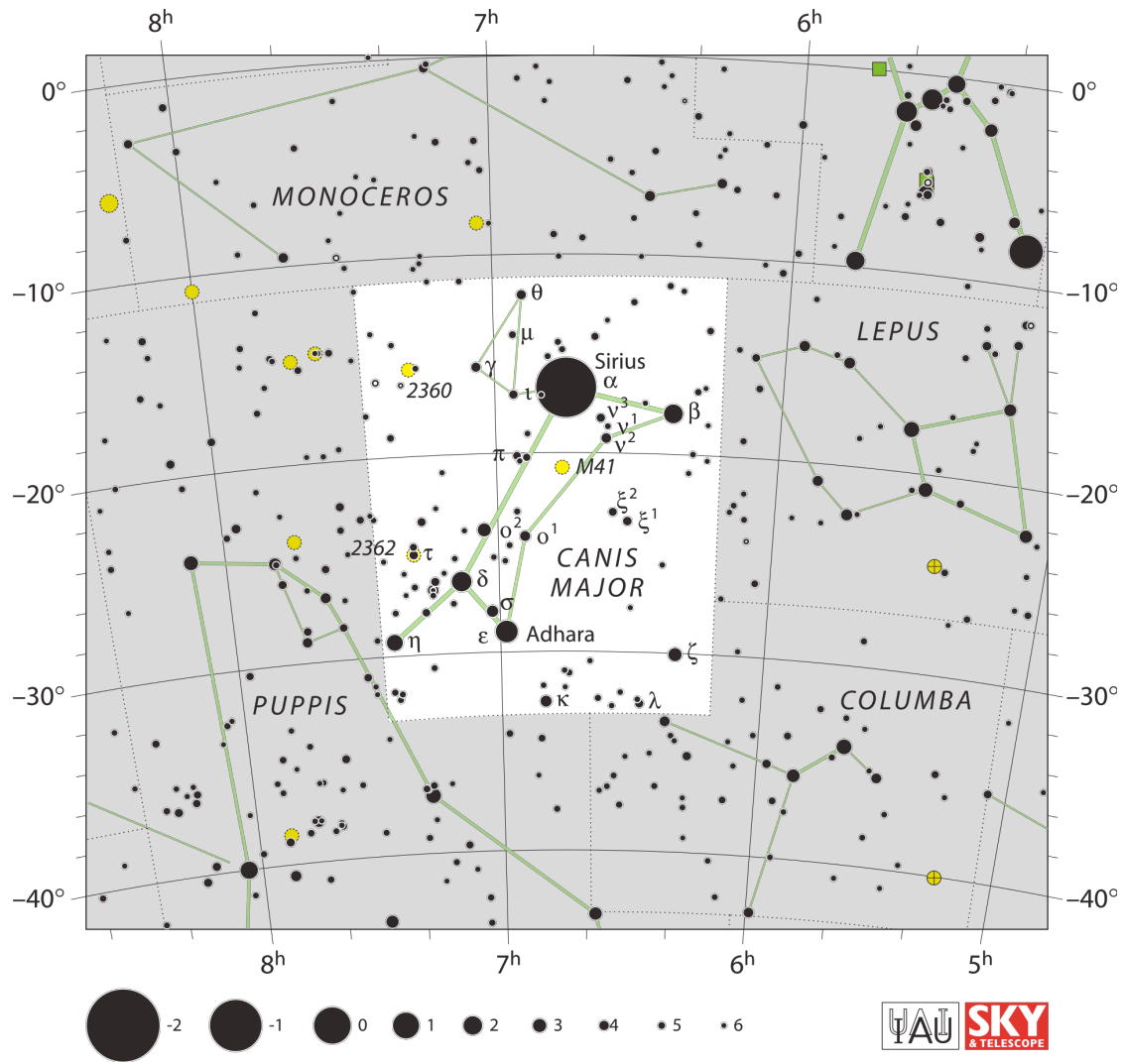

*Figure 3: The constellation Canis Major, featuring the star Sirius. Image: IAU.*

Astrophysical theory tells us that star colour is denoted by surface temperature. But star colours can vary to an observer depending on effects of reddening, Mie scattering, and Rayleigh scattering as the starlight passes through the Earth's atmosphere and turbulent pockets of air. This has a stronger effect for stars low on the horizon. The Inuit near Igloolik say the star Sirius (which is a binary system with a white Class A star and a small white-dwarf star) can change colour between red and white. Sirius never rises more than nine degrees above the horizon when seen from Igloolik. As the white stars' light enters our atmosphere at a low angle, air pockets alter the refractive index of the incoming light. This chromatic aberration causes starlight to change colour when twinkling. Being low on the horizon, Sirius is also affected by reddening, where bluer wavelengths of light are scattered while redder wavelengths pass through to the observer. The rapid shifting of Sirius' light from red to white is probably the reason it is associated with two red-and-white Arctic foxes fighting with each other to gain access to a single foxhole (MacDonald 1998: 75). Historical records suggest the star has appeared ruddy in colour in the past (Ceragioli 1993), and some analyses favours the interpretation of Sirius' changing colours being due to atmospheric extinction when viewed at low-altitudes (Chapman-Rietschi 1995).

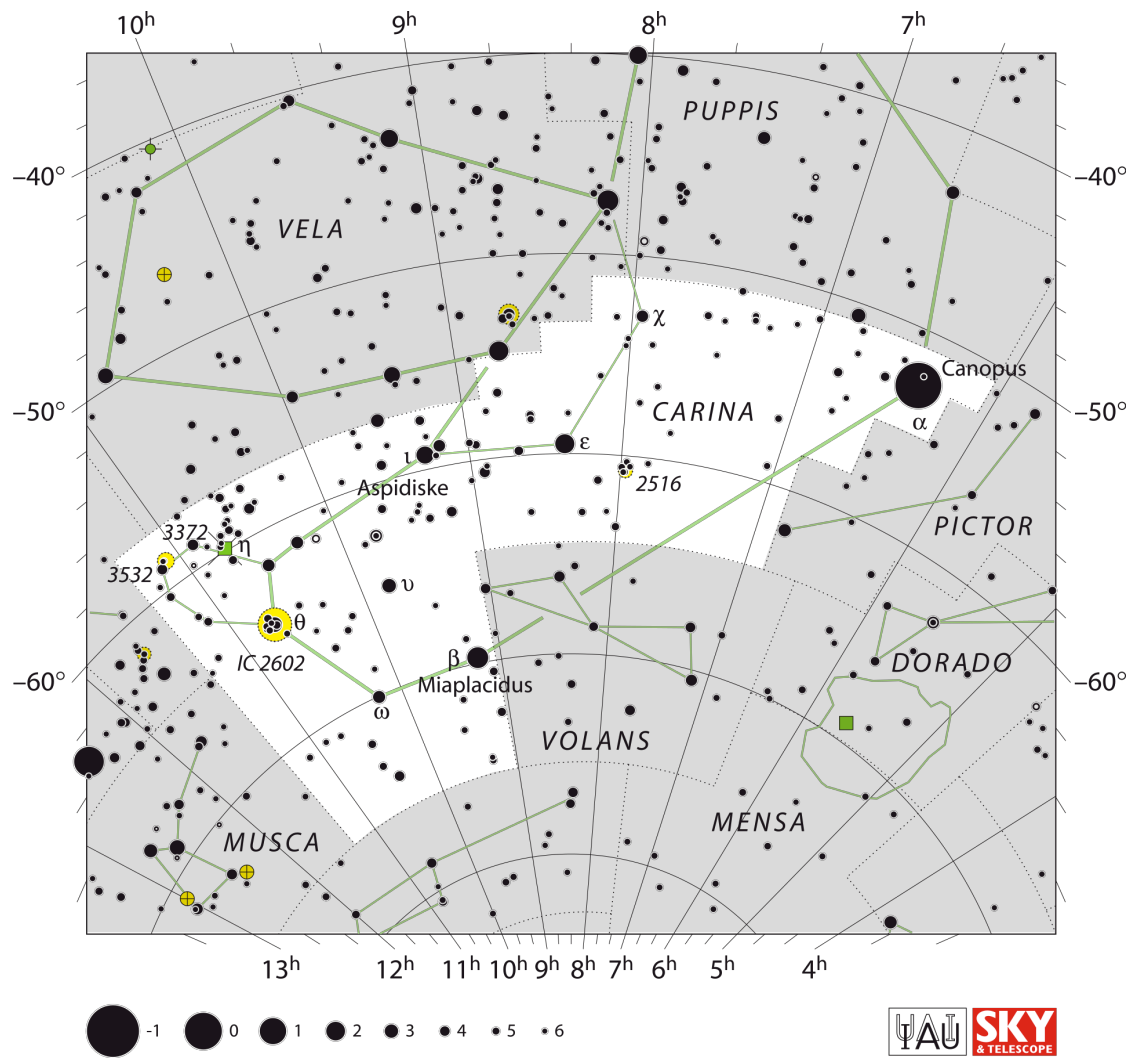

*Figure 4: The constellation Carina, featuring the star Canopus. Image: IAU.*

The water absorption coefficient for visible light is much higher at green and red wavelengths (500-780 nm) but is lowest at blue wavelengths (390-450 nm; Figure 5; Pope & Fry 1997). This can make the Moon and stars in humid conditions appear bluer in colour. Meriam co-author Alo Tapim describes this effect when he stipulates that the stars appearing very blue in colour is a sign of impending rain. Similarly, Meriam co-author John Barsa notes that blue/white stars twinkling rapidly signal approaching storms.

Dust and humidity can also affect the colours of stars visible to an observer. Dust can scatter light particles, making them appear different colours. This is dependent on the size of the particles. If they are larger than the visible wavelengths of light, all wavelengths of light are scattered equally (Mie scattering). If dust particles suspended in the atmosphere measure are significantly smaller than the wavelength of the sunlight or starlight, they can scatter those wavelengths (Tyndall effect). Dust caused from volcanism or bushfires tend to be on the order of 500-800 nm across, scattering red light. If the suspended particles are smaller than 300-400 nm, which can occur when certain aerosols from vegetation interact with ozone, they can scatter blue light. Conditions like this can give the Moon a red or blue tinge, and the same principle applies to starlight. Thus, things like bushfires and volcanic eruptions could cause changes in starlight colour, but the elders did not tie them to weather and seasonal prediction, as they are less common, one-off events.

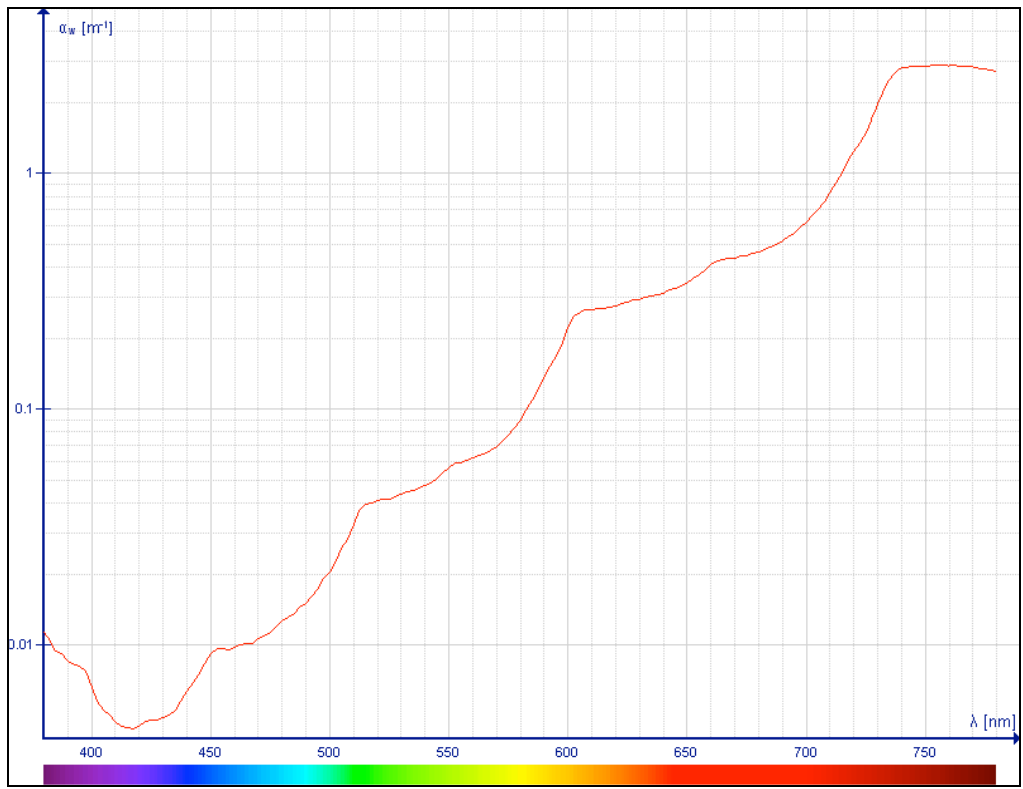

*Figure 5: The absorption coefficient of water in the visible spectrum (using data from Pope & Fry 1997, and Kou et al. 1993). Wikimedia Commons.*

Scintillating stars that appear sharp to the eye indicate dry, clear skies. Humidity and ice crystals in the atmosphere can blur the stars. Thus, the jagged appearance of scintillating stars denotes moisture, thus warning of a coming storm or cyclone. When most of the visible stars are blue, it indicates moisture, as the red and green wavelengths of light are absorbed by the humidity. In the Torres Strait, twinkling blue stars indicate hot weather with increasing humidity as the wet monsoon season approaches. The blurry characteristic of stars during periods of high humidity is known to many farmers, who use the folk saying "When stars become a muddle, Earth becomes a puddle" (Dappen 1989: 30).

Although scintillation and seeing effects alter the apparent properties of stars, they are not well correlated (Aerts *et al.* 2010: 306). This means that observing stellar scintillation may, but will not necessarily, include changes in colour, position, or clarity. This introduces a degree of error and uncertainty in forecasting atmospheric conditions. This is why it is important to note that stars are just one of *many* natural objects observed for predicting weather.

To the trained eye, subtle changes in apparent stellar properties can be used to determine changes in atmospheric conditions. That can be combined with other local data, such as the behavior of animals, clouds, and other natural objects and phenomena as part of a comprehensive weather forecasting system.

**Conclusion**

We see that Indigenous people from different parts of the world utilise stellar scintillation as a seasonal and weather predictor, sharing knowledge foundations with Western science. This research provides insight into the scientific underpinnings of Indigenous Knowledge Systems through careful long-term observations of local climate, geography, and astronomy. This paper is a starting point for future scholarship that explores the role and usage of stellar scintillation and related phenomena in global Indigenous cultures.

**Acknowledgments**

Hamacher acknowledges that the knowledge shared by Indigenous consultants remains their intellectual property. Elders who shared knowledge and were involved in the development of the manuscript were invited to be co-authors on this paper. Three accepted: John Barsa, Segar Passi, and Alo Tapim. Meriam custodians granted approval for the contents of this paper to be published. We dedicate this paper to John Barsa, who passed away in January 2018, prior to the paper's publication.

The authors gratefully acknowledge the Mer Ged Kem Le, Doug Passi, Michael Passi, Martin Nakata, Aven Noah, Leah Lui-Chivizhe, Christopher Cannon, and the anonymous consultants for their contributions and feedback. Hamacher is funded by the Australian Research Council (DE140101600) and interviews with Meriam communities in the Torres Strait were approved by the Monash University Human Research Ethics Committee (HC15035).